\documentclass[sigconf,screen,table]{acmart}

\usepackage{xcolor}
\usepackage[ruled,vlined]{algorithm2e}
\usepackage{verbatimbox}
\usepackage{makecell}
\usepackage{todonotes}
\usepackage{xspace}
\usepackage{enumitem}
\setlist{nosep}
\usepackage{cleveref}
\usepackage{caption}
\usepackage{subcaption}
\usepackage{placeins}
\usepackage{pifont} 
\usepackage{multirow}
\usepackage{arydshln} 

\AtBeginDocument{%
  \providecommand\BibTeX{{%
    \normalfont B\kern-0.5em{\scshape i\kern-0.25em b}\kern-0.8em\TeX}}}

\usepackage{array}
\usepackage{amsmath}
\usepackage{xspace}

\newcommand{\eg}{e.\,g., }
\newcommand{\ie}{i.\,e., }
\newcommand{\wrt}{w.\,r.\,t.\, }

\newcommand{\ndcg}{\text{nDCG}\xspace}

\newcommand{\ndcgten}{\text{nDCG}@\ensuremath{10}\xspace}
\newcommand{\precten}{\text{precision}@\ensuremath{10}\xspace}
\newcommand{\recten}{\text{recall}@\ensuremath{10}\xspace}
\newcommand{\apten}{\text{AP}@\ensuremath{10}\xspace}
\newcommand{\covten}{\text{coverage}@\ensuremath{10}\xspace}

\def\Tabref#1{Table~\ref{#1}}

\newcommand{\mf}{\text{MF}\xspace}
\newcommand{\dmf}{\text{DeepMF}\xspace}
\newcommand{\rand}{\text{Rand}\xspace}
\newcommand{\pop}{\text{Pop}\xspace}

\newcommand{\clcrec}{\text{CLCRec}\xspace}

\newcommand{\mlonem}{\text{ML-1M}\xspace}
\newcommand{\onion}{\text{Onion}\xspace}
\newcommand{\amazonvid}{\text{Amazon}\xspace}

\newcommand{\dropoutnet}{\text{DropoutNet}\xspace}
\newcommand{\dropoutnetBest}{\text{DropoutNet}\textsubscript{\text{best}}\xspace}
\newcommand{\dropoutnetTwo}{\text{DropoutNet}\textsubscript{\text{one}}\xspace}

\newcommand{\sbrec}{\texttt{SiBraR}\xspace}
\newcommand{\sbreclong}{a multimodal \textbf{Si}ngle-\textbf{Bra}nch embedding network for \textbf{R}ecommendation\xspace}
\newcommand{\sbreclongna}{multimodal \textbf{Si}ngle-\textbf{Bra}nch embedding network for \textbf{R}ecommendation\xspace}
\newcommand{\sbrecBest}{\text{SiBraR}\textsubscript{\text{best}}\xspace}
\newcommand{\sbrecTwo}{\text{SiBraR}\textsubscript{\text{one}}\xspace}
\newcommand{\sbrecItem}{\texttt{Item-SiBraR}\xspace}
\newcommand{\sbrecUser}{\texttt{User-SiBraR}\xspace}

\newcommand{\cfbr}{\text{CF}\xspace} 
\newcommand{\cfbrs}{\text{CFs}\xspace} 

\newcommand{\cbr}{\text{CBRS}\xspace} 
\newcommand{\cbrs}{\text{CBRSs}\xspace} 

\newcommand{\baserec}{\text{Base}\xspace} 

\newcommand{\repo}{\href{https://github.com/hcai-mms/SiBraR---Single-Branch-Recommender}{https://github.com/hcai-mms/SiBraR---Single-Branch-Recommender}}

\def\Figref#1{Figure~\ref{#1}}

\def\eqref#1{equation~\ref{#1}}

\def\1{\bm{1}}

\DeclareMathAlphabet{\mathsfit}{\encodingdefault}{\sfdefault}{m}{sl}
\SetMathAlphabet{\mathsfit}{bold}{\encodingdefault}{\sfdefault}{bx}{n}

\makeatletter
\gdef\@copyrightpermission{
\begin{minipage}{0.15\columnwidth}
\href{https://creativecommons.org/licenses/by-nc/4.0/}{\includegraphics[width=0.9\textwidth]{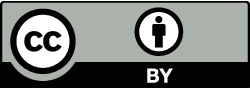}}
\end{minipage}\hfill
\begin{minipage}{0.85\columnwidth}
\href{https://creativecommons.org/licenses/by/4.0/}{This work is licensed under a Creative Commons Attribution International
4.0 License.}
\end{minipage}
\vspace{5pt}
}
\makeatother
\copyrightyear{2024}
\acmYear{2024}
\setcopyright{rightsretained}
\acmConference[RecSys '24]{18th ACM Conference on Recommender Systems}{October 14--18, 2024}{Bari, Italy}
\acmBooktitle{18th ACM Conference on Recommender Systems (RecSys '24), October 14--18, 2024, Bari, Italy}\acmDOI{10.1145/3640457.3688009}
\acmISBN{979-8-4007-0505-2/24/10}

\begin{document}

\title[Multimodal Single-Branch Recommender for Cold-Start and Missing Modality]{A Multimodal Single-Branch Embedding Network for Recommendation in Cold-Start and Missing Modality Scenarios}
%

\newif\ifworkinprogress
\workinprogressfalse

\newcommand{\newrs}{\cellcolor{blue!25}}
\newcommand{\cmark}{\ding{51}}

\ifworkinprogress
	\newcommand{\ms}[1]{\textcolor{blue}{{[Markus] #1}}}
	\newcommand{\mm}[1]{\textcolor{olive}{{[Marta] #1}}}
	\newcommand{\sn}[1]{\textcolor{green}{{[Shah] #1}}}
    \newcommand{\ah}[1]{\textcolor{red}{{[Anna] #1}}}
	\newcommand{\ch}[1]{\textcolor{orange}{{[Christian] #1}}}

    \newcommand{\newtext}[1]{\textcolor{purple}{#1}}    
\else
    \newcommand{\ms}[1]{}
    \newcommand{\mm}[1]{}
    \newcommand{\sn}[1]{}
    \newcommand{\ah}[1]{}
    \newcommand{\ch}[1]{}
    
    \newcommand{\newtext}[1]{#1}
    \newcommand\sout[1]{}
\fi

\author{Christian Ganhör}
\authornotemark[1]
\email{christian.ganhoer@jku.at}
\orcid{0000-0003-1850-2626}
\affiliation{%
  \institution{Institute of Computational Perception, Johannes Kepler University Linz}
  \streetaddress{Altenberger Straße 69}
  \city{Linz}
  \country{Austria}
}

\author{Marta Moscati}
\email{marta.moscati@jku.at}
\orcid{0000-0002-5541-4919}
\affiliation{%
  \institution{Institute of Computational Perception, Johannes Kepler University Linz}
  \streetaddress{Altenberger Straße 69}
  \city{Linz}
  \country{Austria}
}
\authornote{These authors are listed alphabetically and contributed equally to the paper.}

\author{Anna Hausberger}
\email{anna.hausberger@jku.at}
\orcid{0009-0003-5562-681X}
\affiliation{%
  \institution{Institute of Computational Perception, Johannes Kepler University Linz}
  \streetaddress{Altenberger Straße 69}
  \city{Linz}
  \country{Austria}
}

\author{Shah Nawaz}
\email{shah.nawaz@jku.at}
\orcid{0000-0002-7715-4409}
\affiliation{%
  \institution{Institute of Computational Perception, Johannes Kepler University Linz}
  \streetaddress{Altenberger Straße 69}
  \city{Linz}
  \country{Austria}
}

\author{Markus Schedl}
\email{markus.schedl@jku.at}
\orcid{0000-0003-1706-3406}
\affiliation{
  \institution{Institute of Computational Perception, Johannes Kepler University Linz and Human-centered AI Group, AI Lab, Linz Institute of Technology}
  \city{Linz}
  \country{Austria}
}

\renewcommand{\shortauthors}{}

\begin{abstract}
    Most recommender systems adopt collaborative filtering (CF) and provide recommendations based on past collective interactions. Therefore, the performance of CF algorithms degrades when few or no interactions are available, a scenario referred to as \textit{cold-start}. To address this issue, previous work relies on models leveraging both collaborative data and side information on the users or items. Similar to multimodal learning, these models aim at combining collaborative and content representations in a shared embedding space. In this work we propose a novel technique for multimodal recommendation, relying on \sbreclong (\sbrec). Leveraging weight-sharing, \sbrec encodes interaction data as well as multimodal side information using the same single-branch embedding network on different modalities. This makes \sbrec effective in scenarios of missing modality, including cold start. Our extensive experiments on large-scale recommendation datasets from three different recommendation domains (music, movie, and e-commerce) and providing multimodal content information (audio, text, image, labels, and interactions) show that \sbrec significantly outperforms \cfbr as well as state-of-the-art content-based RSs in cold-start scenarios, and is competitive in warm scenarios. We show that \sbrec's recommendations are accurate in missing modality scenarios, and that the model is able to map different modalities to the same region of the shared embedding space, hence reducing the modality gap.

\end{abstract}


\maketitle

\section{Introduction}
    Recommender systems (RSs) offer powerful solutions for navigating the vast amounts of multimedia content available today, helping users find new and interesting items. Modern RSs can handle diverse media types such as text, audio, image, and video, both as their output, \ie items to recommend, and as modalities the items in the catalog are composed of. While the vast majority of RSs still build upon collaborative filtering (\cfbr) techniques, research advancements in multimedia content analysis and neural learning-to-rank models have enabled the development of effective content-based RSs~\cite{10.1145/3407190} (\cbrs). Since \cfbr techniques solely rely on past collective interaction data, these algorithms often struggle to provide accurate recommendations when users or items do not appear in the past interactions, a scenario referred to as \textit{cold start}. On the other hand, \textit{pure} \cbrs are often effective for cold start, but relying solely on content information might limit the performance of such algorithms in warm scenarios. 
    Therefore, the most effective way to provide accurate recommendations both in  warm and cold-start scenarios is to simultaneously leverage collaborative data and side information on items and users. The resulting hybrid architectures\footnote{We refer to hybrid RSs as \cbrs and to RSs leveraging only content information as \textit{pure} \cbrs.} therefore naturally belong to the domain of \textit{multimodal learning}, where one or more content modalities have to be used in combination with 
    interaction data. As such, \cbrs are still prone to decreases in performance when one or more input modalities are missing. To address this issue, we propose the use of \sbreclong (\sbrec, pronounced ``zebra''
    ). 
    \sbrec leverages a single-branch network architecture coupled with weight sharing to embed multiple modalities. This allows mapping different modalities of the same entity (\ie a user or an item) to similar positions in the shared embedding space, therefore mitigating scenarios in which one or more modalities are missing. This way of addressing the missing modality scenario naturally results in an improvement in performance in cold-start scenarios with respect to both \cfbr approaches and other \cbrs. The contributions of this paper can be summarized as follows:
    \begin{itemize}[leftmargin=*]
        \item We propose \sbrec, a novel \cbr which leverages multimodal information
        for effective recommendations in standard as well as cold-start and missing-modality scenarios. 
        \item We perform extensive quantitative experiments to assess the accuracy of \sbrec's recommendations, and compare it to other traditional and state-of-the-art methods.
        \item We analyze the impact of missing modalities on the performance of \sbrec.
        \item We investigate the shared embedding space of \sbrec, showing that the model  maps different modalities to the same region of the shared embedding space, reducing the modality gap.
    \end{itemize}
    
    In the remained of this paper, we first review work related to ours (Section~\ref{sec:related_work}). We then introduce the notation and describe our method (Section~\ref{sec:methodology}), as well as the experiment setup (Section~\ref{sec:experiments}). We discuss the evaluation results, the impact of missing modalities, and the modality gap (Section~\ref{sec:results}), as well as \newtext{limitations and} possible future extensions of the current work (Section~\ref{sec:conclusions}).

\section{Related Work}
    \label{sec:related_work}
    Relevant previous work falls within two strands of research: \cbrs 
    for targeting cold start
    and multimodal representation learning. \newtext{As our main focus are cold start and missing modality, we exclude general \cbrs not designed for these scenarios and refer the reader to Musto et al.~\cite{musto_rshandbook}.}
    \subsection{Content-Based Cold-start Recommendation}
        Recommending items in cold-start scenarios is one of the main challenges of RSs~\cite{rshandbook, mrs};  For a recent literature review we refer the reader to Panda et al.~\cite{panda_cold_start_review}.  Recent approaches~\cite{megnn, metarecsys, metaedl, paml, melo, wdof, tdas} use meta-learning for cold-start recommendation, which might not tackle scenarios where no interactions at all are available for certain users or items. 
        When certain users or items completely lack user--item interactions, RSs often rely on side  
        information
        . This is done by either adding a content loss term, or adapting the learning process without any additional loss term~\cite{dropoutnet}; Models of the first type are often referred to as \textit{explicit}
        , and those in the second as \textit{implicit}
        ~\cite{ood_cold_start}
        .\footnote{Although the term implicit is used for both  \cbrs strategies and the type of user--item interaction data, its meaning
        should be clear from the context.} Within explicit methods, 
        Wei et al.~\cite{clcrec}
        combine \cfbr
        with contrastive losses between two representations of items, one obtained from collaborative signals and the other from content representations; The resulting deep neural network (NN)-based content encoder is therefore trained to obtain representations that are as similar as possible to the representations of the collaborative signals, and that can be used in their absence
        . Similarly, Li et al.~\cite{li2024nemcf} use contrastive learning losses in graph-based RSs. The contrastive loss aims at encoding content and interaction information using as positive pairs the pairs of items often co-occurring for the same users. Their approach shows an improvement in performance in recommendations for e-commerce. Wu et al.~\cite{uima} propose the use of three loss terms to address cold start with content: reconstruction loss terms on two autoencoder architectures, one for item content and the other for user interaction data, and a \cfbr loss. Wang et al.~\cite{ood_cold_start} address the problem that due to the development of content production over time, recently added items without interactions might have a different distribution of content features with respect to ``older'' items, and might therefore be underrepresented in recommendations.
        To solve this issue, they use a sample interpolation strategy and loss functions that align the representations of interpolations of items with the interpolation of the item representations.  Barkan et al.~\cite{cb2cf} and Zhu et al.~\cite{zhu2019heater} propose the use of mean square error as additional loss function to minimize the distance between \cfbr and content representations. \newtext{Wang et al.~\cite{wang2018lrmm} propose the use of an encoder-decoder architecture taking as input the concatenation of multimodal data of both users and items; The autoencoder is optimized to simultaneously reconstruct missing modalities and predict the user--item interactions, such as the ratings.} Cao et al.~\cite{pre_training_cold_start} separately train a graph NN to model item content similarities, and a transformer-based RS, and propose the use of Euclidean distance as loss term  to align the resulting representations. \newtext{Similarly, Zhang et al.~\cite{zhang2021lattice} model item content similarities as item embeddings representing a modality-aware content graph; The resulting embeddings are added to the item embeddings obtained from established CF algorithms and the result is used to obtain the final recommendation score.} Gong et al.~\cite{gong2023fidr} augment an interaction-based graph by constructing representations of cold items based solely on content. Pulis et al.~\cite{pulis2023snn} propose a query-by-multiple-example~\cite{bogdanov2010musiccontent} music RS based on item representations extracted from the audio signal of music tracks in a way to maximize the similarity of tracks of the same music genre. Shalaby et al.~\cite{shalaby2022m2trec} apply transformers~\cite{vaswani2017transformers_attention} to cold-start session-based RSs and include loss terms to predict both the next item and its category.
        Magron et al.~\cite{mrs_content_cold} propose two extensions of neural collaborative filtering~\cite{he2017ncf} with the inclusion of a NN that accepts item content as input. In the first variant, which falls into the category of explicit models, the NN is optimized to minimize the Euclidean distance between its output and the item collaborative representation. In the second variant, falling into the category of implicit models, the deep content feature extractor directly predicts the item collaborative embedding. The category of implicit cold-start methods includes the work of Raziperchikolaei et al.~\cite{item_cold_start_nn}, who propose the use of item content as input to a NN and to further use the resulting representations as weights in the first layer of an encoder for the user interaction profiles.
        Cai et al.~\cite{ihgnn} and Behar et al.~\cite{recruitergcn} model collaborative as well as content information on heterogeneous graphs to learn embeddings that are representative of both. R et al.~\cite{feature_concat} use NNs to obtain latent representations of users, items, and their side information; these are further concatenated and fed to a NN to obtain the score of a user--item pair. The use of concatenation of content and interaction representations is also proposed by Volkovs et al.~\cite{dropoutnet}, who further apply dropout as regularization technique to mimic the lack of information during training, showing that this leads to a better performance in cold-start scenarios. Gong et al.~\cite{llm_cikm} leverage large-language models (LLMs) to obtain domain-invariant item representations, demonstrating their effectiveness in cold-start scenarios, while Sanner et al.~\cite{sanner2023llmrec} leverage pre-trained LLMs to provide recommendations from both item-based and language-based preferences, demonstrating their better performance \wrt standard \cfbr in the near cold start (\ie when some user or items have few interactions). The novelty of our work relies on the use of a single-branch network to encode multimodal content information as well as collaborative data. Since our proposed \sbrec can be leveraged both with and without the use of a content loss, it does not immediately fall into either categories of explicit or implicit \cbrs, hence establishing a new paradigm.

    \subsection{Multimodal Representation Learning}
        Multimodal learning leverages information from multiple modalities, such as audio, image, or text, to improve the performance on various machine learning (ML) tasks, such as classification, retrieval, or verification~\cite{multimodal_survey,xu2023multimodal}. 
        Although multimodal tasks differ, the way they are usually addressed is very similar and relies on learning joint representations from multiple modalities. 
        Several multimodal methods explored the use of NNs to map multimodal information to joint representations.
        For example, multi-branch NNs use separate independent and modality-specific NNs to map each modality to a joint embedding space~\cite{reed2016learning, nagrani2018seeing,nagrani2018learnable,saeed2022fusion,arshad2019aiding}. Recently, multi-branch architectures have also been extended with the use of transformers ~\cite{vaswani2017transformers_attention,lu2019vilbert,tan2019lxmert,xu2023multimodal}.
        Such multi-branch methods have achieved remarkable performance using modality-complete information, \ie when all modalities are available for training and evaluation. 
        However, they suffer from performance deterioration if a modality is missing either during training or evaluation~\cite{ma2021smil,lin2023missmodal,lee2023multimodal}.
        Recently, single-branch models~\cite{saeed2023single}, \ie models that share the same embedding NN across multiple modalities, have shown promising results in multimodal learning. Although effective for other multimodal ML tasks, these models have never been translated to the domain of recommendation. Our work fills this gap in the current status of research by proposing \sbrec, a novel multimodal RS based on a single-branch architecture. Furthermore, we analyze the effectiveness of single-branch architectures in missing modality scenarios, which are related to cold start in recommendation.         
        
\section{Methodology}
    \label{sec:methodology}
    In this section, we introduce our \sbreclongna 
    (\textit{\sbrec}). We introduce the notation and mathematical formulation of the problem. We then describe how \sbrec leverages multimodal information for recommendation in warm and cold-start scenarios. \\
    \textit{Notation.}
        We denote with $\mathcal{U} = \{u_i\}_{i=1}^{M}$ the set of $M$ users and $\mathcal{I} = \{i_j\}_{j=1}^{N}$ the set of $N$ items and refer to users and items as \textit{entities}. We consider the scenario of implicit feedback and represent the user--item interactions as a binary matrix in $R\in \mathbb{R}^{M\times N}$ with nonzero entries $R_{ij} = 1$ for a positive interaction of user $u_i$ with item $i_j$. We refer to rows of $R$ as user profiles $\mathbf{u}_i\in \mathbb{R}^{N}$; Profiles store information on the items with which user $u_i$ interacted. Analogously, we refer to columns of $R$ as item profiles $\mathbf{i}_j\in \mathbb{R}^M$. Cold-start users (items) are those without interactions, \ie with an empty profile. 
        Each user $u_i$ is represented by a set of \textit{modalities} $\mathcal{M}^{\text{user}}_{i}$, consisting of \textit{feature vectors} representing available side information, \eg gender and country, and their profile if not empty. Empty profiles ($\mathbf{u}_i = \mathbf{0}$) of cold-start users are not included in the set of modalities available for the user and are treated as missing modalities. Analogously, each item $i_j$ is represented by a set of item modalities $\mathcal{M}^{\text{item}}_j$, consisting of feature vectors representing the item's available side information, \eg image or text data, and their profile if not empty. \\
    \textit{\sbrec.}
        The proposed model outputs a recommendation score~$\hat{y}_{ij}$ for a given user--item 
        pair $(u_i, i_j)$, by assigning an embedding vector~$\mathbf{e}_{i}$ to user~$u_i$ and an embedding vector $\mathbf{e}_{j}$ to item~$i_j$, and by computing their scalar product $\hat{y}_{ij} = \mathbf{e}_{i}\cdot \mathbf{e}_{j}$. The ordered list of the top~$k$ items that were not already interacted with by user~$u_i$ (\ie such that $R_{ij}=0$) constitutes the recommendations for user~$u_i$. 
        Based on the assumption that different modalities of the same entity contain similar semantic representations, \sbrec aims at constructing
        embedding vectors for accurate recommendations by taking any available modality as input and projecting them into the same shared space.
        Therefore, \sbrec leverages weight-sharing and uses the same deep NN~$g$ to embed different modalities. The network~$g$ is optimized to provide accurate recommendations with any of the modalities as input. This encourages the model to encode modalities of the same entity to similar embeddings; For instance, for the same user~$u_i$:  $g(m^i_1)\simeq g(m^i_2) \; \forall m_1, m_2 \in \mathcal{M}^{\text{user}}_{i}$.     
        \begin{figure}
    \centering
    \includegraphics[width=\columnwidth]{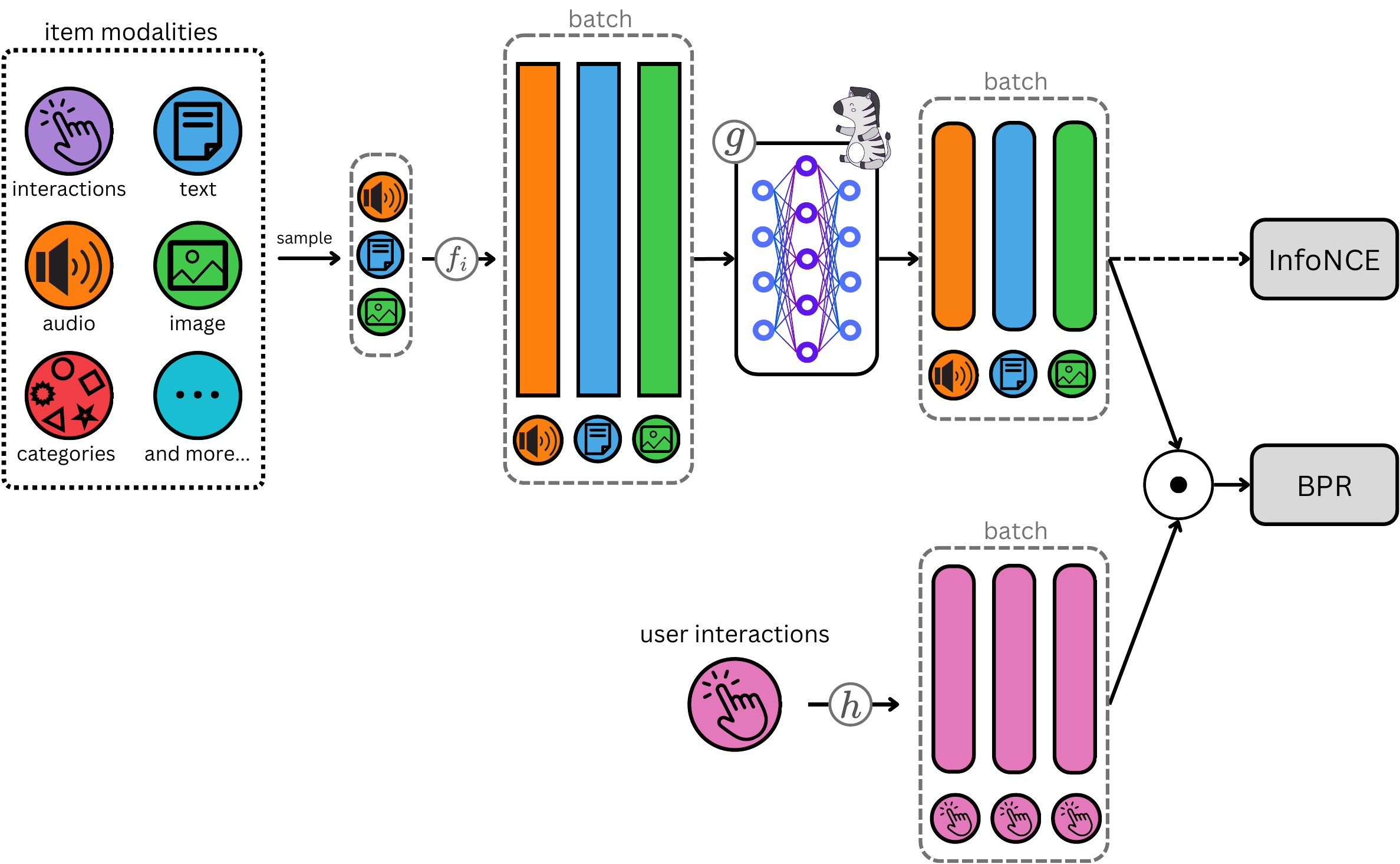}
    \caption{\sbrecItem model and training procedure. The \sbrec network represents the single-branch encoding network $g$ shared across modalities. For each user--item interaction pair $(u_i, i_j)$ in the training set, the recommendation loss $\mathcal{L}_\text{BPR}$ is computed between positive and negative items. The contrastive loss $\mathcal{L}_\text{SInfoNCE}$ is computed for two item modalities and for the set of items consisting of positive item and set of negatives.}
    \label{fig:itemmodel}
\end{figure}
        We now describe our proposed \sbrec architecture in its variant leveraging multimodal item side information. We denote this variant by \sbrecItem. \sbrecItem is designed to tackle scenarios of item cold start and missing item modality, while the \sbrec variant leveraging multimodal user side information \sbrecUser is designed for user cold start and missing user modality. \\
    \textit{\sbrecItem.}
        \Figref{fig:itemmodel} sketches the architecture and training strategy of \sbrecItem, while Algorithm~\ref{alg:batchloss} provides the pseudocode for the computation of the batch loss used for its training. In the case of \sbrecItem, we denote the item embedding function with $g$ and the user embedding function with $h$. For \sbrecItem, the item embedding function $g$ is the core component, and is described in detail below, while for the user embedding function $h$ we consider either an embedding lookup table, where for each user a unique embedding is randomly initialized and optimized during training, or an architecture similar to DeepMF~\cite{dmf}, employing deep NNs to embed the users by taking their profile as input.\footnote{\newtext{In preliminary experiments we compared versions of SiBraR with different underlying RS architectures, including variants similar to DeepMF, \ie applying a NN to the user and item interaction profile, or similar to MF, \ie applying an embedding layer on the IDs of users and items. We selected the architecture reaching the highest accuracy of recommendations on the validation set, for the subsequent experiments.}}  During training, \sbrecItem samples a set $\mathcal{N}_{ij}$ of $n_\text{neg} = |\mathcal{N}_{ij}|$ negative items uniformly at random. For each user--item pair, the user embedding is obtained as $\mathbf{e_{i}} = h(u_i)$. The model then samples $n_\text{mod}$ modalities uniformly at random from the set of item modalities $\mathcal{M}^\text{item}$. The positive item and each of the negative items are embedded as follows. First, each sampled modality is projected to the input dimension of the shared network $g$  with a shallow network  $f_i$ consisting of a linear layer  and an activation function. The resulting vectors are then passed through the single-branch network $g$, \ie the same network is used irrespective of the modality. The resulting vectors are then averaged to compute the embedding of the item to be used to compute the recommendation score:
        \begin{equation}
            \mathbf{e}_{j} = \frac{1}{n_\text{mod}} \left( g(m_1) + \dots + g(m_{n_\text{mod}})\right)
        \end{equation}
        The user and item embeddings, $\mathbf{e_i}$ and $\mathbf{e_j}$, respectively, are used to compute the recommendation scores by taking their scalar product. This is done for both the positive item $i_j$, for which $\hat{y}_{ij} = \mathbf{e}_{i}\cdot \mathbf{e}_{j}$, and  for the negative items  $i_k \in \mathcal{N}_{ij}$, for which $\hat{y}_{ik} = \mathbf{e}_{i}\cdot \mathbf{e}_{k}$. Based on the scores, the recommendation loss for Bayesian Personalized Ranking (BPR)~\cite{bpr} is computed. Therefore, for a single user--item interaction pair $(u_i, i_j)$, the recommendation loss is given by 
        \begin{equation}
            \mathcal{L}^{(u_i, i_j)}_\text{BPR} = 
            \sum_{k\in \mathcal{N}_{ij}} 
            \ln{\sigma(\hat{y}_{ij} - \hat{y}_{ik})}.
            \label{eq:bpr}
        \end{equation}
        
        While sharing the embedding network has been proven effective to combine modalities in multimodal ML~\cite{saeed2023single}, we also consider the use of a contrastive loss to further align the embeddings from different modalities. When the contrastive loss is not applied, $n_\text{mod}$ is set to 1 and only one modality is sampled for each user--item interaction pair $(u_i, i_j)$. As the modality sampling is done per interaction pair, over the course of training, all available modalities are used. \footnote{Note that different user--item interaction pairs may correspond to different sampled modalities.} %
        When the contrastive component is applied, $n_\text{mod}$ is set to $2$ and in addition to the recommendation loss, \sbrecItem models apply the symmetric Info Noise Contrastive Estimation  loss~\cite{vandenoord2018symmetricinfonce,radford2021clip} $\mathcal{L}_\text{SInfoNCE}$ between the two item modality embeddings. To this purpose, we contrast the two modalities of the positive item $i_j$ with the two modalities of the negative samples in $\mathcal{N}_{ij}$. The loss between modalities $m_1$ and $m_2$ is therefore computed as $\mathcal{L}_\text{SInfoNCE} = \mathcal{L}^{12}_\text{InfoNCE} + \mathcal{L}^{21}_\text{InfoNCE}$, where $\mathcal{L}^{12}_\text{InfoNCE}$ is the InfoNCE~\cite{vandenoord2018symmetricinfonce} contrastive loss between modalities $m_1$ and $m_2$: 
        \begin{equation}
            \mathcal{L}^{12}_\text{InfoNCE} =  - \sum_{(u_i, i_j)}\ln{\frac{\mathrm{e}^{(g(m^{j}_1) \cdot g(m^{j}_2))/\tau}}{\sum_{k \in \{j\}\cup \mathcal{N}_{ij}}\mathrm{e}^{(g(m^{j}_1) \cdot g(m^{k}_2))/\tau}}},
        \end{equation}    
        and $\tau$ represents the temperature parameter, considered as a hyperparameter as described in Section~\ref{sec:experiments}. Intuitively, $\mathcal{L}_\text{SInfoNCE}$ aims at penalizing highly dissimilar embeddings of the same items, as well as similar embeddings of different items.
        \input{algorithms/itemsbrec}
        During inference, for each item $i_j\in \mathcal{I}$, \sbrecItem uses all its available modalities $\mathcal{M}^{\text{item}}_j$, encoding them with the single-branch network $g$ and taking the average of the resulting encoding:
        \begin{equation}
            \mathbf{e}_{j} = \frac{1}{\left|\mathcal{M}^{\text{item}}_j\right|} \left( g(m_1) + \dots + g\left(m_{\left|\mathcal{M}^{\text{item}}_j\right|}\right)\right).
        \end{equation}
        Therefore, the embeddings of cold-start items are obtained from their content modalities only, since their profile is treated as missing modality. \\ 
    \textit{\sbrecUser.} 
        The training and inference for \sbrecUser follows a very similar approach, employing an embedding layer or a simple NN for the item, and a single-branch network shared across modalities for the user.
    
\section{Experimental Setup}
    \label{sec:experiments}
    In this section, we describe the experimental setup for our experiments, including 
    the datasets, the baselines used for comparison, the evaluation protocol, the training procedure and hyperparameter tuning. We also provide the code used to carry out our experiments.\footnote{\repo} 
    \subsection{Datasets}
        \label{sec:dataset}
        \begin{center}
\begin{table*}[!htb]\scalebox{0.8375}{
    \begin{tabular}{l|rrrr||cccc|cccc}
         &  &  &  &  &\multicolumn{4}{c|}{User Features} & \multicolumn{4}{c}{Item Features} \\
            & \# users & \# items & \# inter.  & sparsity &Age (\textbf{d}) & Gender (\textbf{c}) & Country (\textbf{c})&Occupation (\textbf{c})&Text (\textbf{v})&Audio (\textbf{v})&Image (\textbf{v})&Genre (\textbf{m})\\ 
        \midrule
        \onion~\cite{music4all_onion} & 5{,}192    & 13{,}610    & 407{,}653         & 0.99423  &--&2&152&--&768&50; 100; 4{,}800&4{,}096&853\\
        \mlonem~\cite{movielens} & 5{,}816    & 3{,}299    & 813{,}792         & 0.95759 &7&2&--&21&768&--&--&18\\
        \amazonvid~\cite{hou2024amazonreviews} & 11{,}454& 4{,}177&87{,}098&0.99818&--&--&--&--&768; 768&--&2{,}048&--\\
    \end{tabular}}
    \caption{Summary of datasets. Left side: Characteristics after pre-processing. Right side: Representation type (\textbf{c}ategorical, \textbf{d}iscrete, \textbf{v}ector, \textbf{m}ultilabel), dimensionality (for \textbf{v}) or number of options (for \textbf{c}, \textbf{d}, and \textbf{m}) of each content feature available. Dimensionalities separated by a colon indicate that more than one feature is available, -- denotes features that are not available or have been neglected (see Section~\ref{sec:dataset}).}
    \label{tab:dataset-overview}
\end{table*}
\end{center}
        We carry out experiments on datasets from three domains: music, movie, and e-commerce. The datasets were selected by considering their popularity for benchmarking RSs, their number of content features, and the multimodality of side information available. We consider an implicit feedback scenario with a binary user--item interaction matrix $R$: $R_{ij}=1$ if user $u_i$ interacted with item $i_j$ and $R_{ij}=0$ otherwise. Table~\ref{tab:dataset-overview} summarizes the characteristics of the datasets after pre-processing (see below) as well as the modalities and dimensionalities of the side information used in our experiments. \\
        \textit{Music4All-Onion} (\onion)
            \footnote{\url{https://zenodo.org/records/6609677}}~\cite{music4all_onion} This large-scale dataset provides user--item interactions, side information on the users and on the items, including several representations of the items content, related to the audio signal, to the lyrics, and to the videoclips. We consider gender and country as user side information. We include representations of the audio signals in terms of i-vectors~\cite{eghbal2015vectors} of dimension $256$ and visual representations of the YouTube videoclips of the music tracks  obtained with a pre-trained instance of ResNet~\cite{he2016resnet}, both as provided by Moscati et al.~\cite{music4all_onion}.
            Additionally, we extend the set of features provided by Moscati et al. with more recent NN architectures from the domains of music information retrieval and natural language processing. More specifically, we encode the $30s$ audio snippets provided by the Music4All dataset~\cite{music4all} with the NNs MusicNN\footnote{\url{https://github.com/jordipons/musicnn/tree/master}}~\cite{pons2018musicnn1, pons2019musicnn2} and Jukebox\footnote{\url{https://github.com/openai/jukebox/}}~\cite{openai2020jukebox}, and the lyrics with the \texttt{all-mpnet-v2}\footnote{\url{https://huggingface.co/sentence-transformers/all-mpnet-base-v2}} pre-trained instance of the SentenceTransformer model~\cite{reimers2019sentencebert} provided by HuggingFace~\cite{wolf2020huggingface}, after applying the same lyrics-specific pre-processing described by Moscati et al.~\cite{music4all_onion}. We restrict to users and items for which all side and content information is available and to users of age between 10 and 80. 
            Following standard practice in the music RS domain~\cite{melchiorre2022protomf, melchiorre2021genderfair}, we restrict to the set of listening events over one year, selecting 2018, as this year was not affected by the outbreak of Covid-19, which impacted the listening behavior of users of music streaming platforms~\cite{liu2020covid, ghaffari2024covidmusic}. Finally, as commonly done in domains of implicit feedback such as music recommendation~\cite{melchiorre2022protomf, melchiorre2021genderfair}, we convert the interactions to binary implicit feedback with a threshold of 2 on the interaction counts (reducing false-positive interactions) and perform 5-core filtering for users and items. \\
        \textit{MovieLens 1M} (\mlonem)
            \footnote{\url{https://grouplens.org/datasets/movielens/1m/}}~\cite{movielens} This dataset consists of movie ratings collected from the MovieLens website.\footnote{\url{https://movielens.org/}} In addition to the user--item interactions, the dataset provides side information on users (age, gender, and occupation) and items (movie genre). Additionally, we crawl Wikipedia for movie plots and encode the product titles with the \texttt{all-mpnet-v2}\footnote{\url{https://huggingface.co/sentence-transformers/all-mpnet-base-v2}} pre-trained instance of the SentenceTransformer model~\cite{reimers2019sentencebert} provided by HuggingFace~\cite{wolf2020huggingface}. We restrict to users and items for which all features are available and to users of age between 10 and 80 years. We apply 5-core filtering and convert the ratings to binary implicit feedback by ignoring the actual values of the ratings, keeping only the interaction information. \\
        \textit{Amazon Video Games} (\amazonvid)
            \footnote{\url{https://amazon-reviews-2023.github.io/index.html}}~\cite{hou2024amazonreviews}
            Amazon Reviews'23 is a large-scale dataset consisting of Amazon Reviews posted from May 1996 to September 2023; This recently released dataset extends the well-established Amazon Reviews'18 dataset~\cite{ni2019amazonreviews}. \amazonvid includes several item features such as title, description, price, and one or more links to images of the products. After considering several categories, we chose to restrict to the category \textit{Video Games}, as this resulted in sets of users, items, and user--item interactions that are large enough to mimic a real-world scenario, and provides characteristics that are different from the two other datasets considered. We restrict to reviews posted between January 2016 and December 2019. \footnote{\newtext{Although the Covid-19 virus was identified in December 2019, the virus spread to other areas of Asia, and then worldwide in early 2020. Therefore, this selection also excludes interactions affected by the outbreak of Covid-19.}}    
            We only consider items that include titles, descriptions, and at least one image of the product in high resolution. We convert the ratings to binary implicit feedback with a threshold of 1, \ie ignoring the actual ratings. We apply 5-core filtering, restricting to users that rated at least 5 different items and to items that have been rated by at least 5 different users. We encode the product titles and descriptions with the \texttt{all-mpnet-v2}\footnote{\url{https://huggingface.co/sentence-transformers/all-mpnet-base-v2}} pre-trained instance of the SentenceTransformer model~\cite{reimers2019sentencebert} provided by HuggingFace~\cite{wolf2020huggingface}. We download the first high-resolution image of the product. Following the same approach as Saaed et al.~in~\cite{saeed2023single}, we resize the image such that the smaller edge has size $256$, crop it at the center to a square of size $224$ pixels, and normalize each channel to the mean and standard deviation of ImageNet~\cite{deng2009imagenet}. We encode the resulting image with the pre-trained instance of \texttt{ResNet-101}~\cite{he2016resnet} provided by TorchVision\footnote{\url{https://pytorch.org/vision/main/models/generated/torchvision.models.resnet101.html}}~\cite{marcel2010torchvision}.

    \subsection{Dataset Splitting}\label{sec:data-split}   
        To evaluate the RSs in warm as well as user and item cold-start scenarios, we split the datasets in three different ways. \\\textit{Warm split.} For every user, we split the data into a training, a validation, and a test set, respectively consisting of $80\%$, $10\%$, and $10\%$ of the number of their interactions, randomly selected. \\  
        \textit{User Cold start.} We split the users into disjoint sets of training,  validation, and test users, respectively consisting of $80\%$, $10\%$, and $10\%$ of the number of users. All interactions of train users are considered during training, and are neglected during validation and testing. All interactions of validation (test) users are considered during validation (testing), and are neglected during training and testing (validation). 
        Since test users are not seen during training,  this scenario is apt to measure the performance of RSs in the user cold-start scenario. \\
        \textit{Item Cold start.} Similar to user cold start, we split the items into disjoint sets of training, validation, and test items, respectively consisting of $80\%$, $10\%$, and $10\%$ of the number of items. All interactions of train items are considered during training, and are neglected during validation and testing. All interactions of validation (test) items are considered during validation (testing), and are neglected during training and testing (validation).
        Since test items are not seen during training, this scenario is apt to measure the performance of RSs in the item cold-start scenario.
        
    \subsection{Baselines}
        We compare the performance of \sbrec with two \cbrs for cold-start recommendation. We also include two \cfbr models leveraging only collaborative data. For \cbr we evaluate \sbrec against \clcrec~\cite{clcrec} and \dropoutnet~\cite{dropoutnet}, selected since they are often used as baselines for \cbr addressing cold-start scenarios, and since they cover both explicit (\clcrec) and implicit (\dropoutnet) approaches.\footnote{See Section~\ref{sec:related_work} for a detailed description of explicit and implicit \cbrs for cold start.} 
        For \cfbr, we include matrix factorization (\mf) with BPR~\cite{bpr} and Deep Matrix Factorization (\dmf)~\cite{dmf}. We select \mf since it is the simplest yet effective latent-representation-based model for recommendation, and \dmf due to its effectiveness in leveraging a two-tower deep NN for recommendation by taking the user and item profiles as inputs. Additionally, we include two naive baselines: \rand, which randomly selects the items to recommend to each user, and \pop, which recommends the same most popular items to all users, measuring popularity in terms of number of interactions in the training set.
        
    \subsection{Metrics}
        \label{sec:evaluation}
        We evaluate the performance of the algorithms on  lists of $k=10$ recommended items, as common in the recommendation domain. We measure accuracy in terms of Normalized Discounted Cumulative Gain (\ndcg)\footnote{We refer the reader to the Additional Material \newtext{available in the code repository} for an evaluation on mean average precision, recall, and catalog coverage.}
        averaged over users in the test set. 
        We test the significance of the best performing algorithm with respect to the others using multiple paired $t$-tests using Bonferroni correction to account for the multiple comparisons. We consider an improvement significant if $p < 0.05$.  During evaluation, for each user in the validation (test) set, we rank all items in the validation (test) set and compute the metrics on the resulting ranking.
        
    \subsection{Hyperparameter Tuning}\label{sec:hyperopt}
        We carry out an extensive hyperparameter optimization to rigorously evaluate the effectiveness of \sbrec and of the baselines. For all models, we tune the learning rate, the weight decay, the embedding dimension, and~--~whenever the model employs NNs~--the number of layers and the number of nodes of each layer. 
        For \cbrs, we treat the training modalities as hyperparameters.\footnote{\newtext{We would like to emphasize that we also investigated combinations of training modalities that did not include the interaction profile as input to the single-branch network. 
        Therefore, interactions are not used as a proxy or anchor for other modalities, as done \eg in ImageBind~\cite{girdhar2023imagebind}. This equal treatment of the modalities in the training phases of our experiments motivates our analogy between cold-start and missing-modality scenarios.}} For \dropoutnet and \sbrec, which support the use of one or more modalities in addition to interaction data, we consider all possible modality combinations of length $1$ up to the number of available modalities. For \clcrec, which only supports the use of one modality at a time in addition to interaction data, we consider all of them separately. For a fair comparison of \clcrec with \dropoutnet and \sbrec, which support the use of more than one modality, we separately select the best \dropoutnet and \sbrec models leveraging only one, or one or more modalities. The variants leveraging one modality are denoted with \dropoutnetTwo and \sbrecTwo, those leveraging one or more with \dropoutnetBest and \sbrecBest. The algorithms reaching the best evaluation metric on the validation set used for the hyperparameter optimization are again evaluated on the test set. The \ndcgten computed on the test set is the one reported for comparison in Section~\ref{sec:results}.\footnote{While we select \dropoutnetBest and \dropoutnetTwo from the sets that also comprise \dropoutnetTwo and \sbrecTwo, since the optimization and model selection is done on the validation set and the final evaluation on the test set, it may occur that \eg \sbrecTwo is better than \sbrecBest on the test set.} For \clcrec and \sbrec we further tune the weight and the temperature of the regularization loss.\footnote{\newtext{We also carried out a comparison of the performance of \sbrec with and without contrastive loss and an ablation study of the impact of the weight and temperature on recommendation accuracy. Due to the page limit, we refer the reader to the repository for the results of these analyses.
}} The hyperparameters are selected through Bayesian optimization, relying on the module Sweep of Weights and Biases platform.\footnote{\url{https://wandb.ai/}} For an overview of the hyperparameters and their value ranges, we refer the reader to the configuration files available in the repository.\footnote{\repo} 
        We set the number of negative samples required during training to $10$. We use \ndcgten on the validation set as validation metric and run all experiments for a maximum of 50 epochs, selecting the model weights that reach the maximum validation metric. We arrest training at an earlier epoch if no improvement on the validation metric is observed for $5$ consecutive epochs.

\section{Results}
    \label{sec:results}
    \begin{center}
\begin{table*}[!htb]
    \centering
    \begin{tabular}{ll|rrr|rrr|rr}
          & & \multicolumn{3}{c|}{Warm} & \multicolumn{5}{c}{Cold} \\
          & & & & & \multicolumn{3}{c|}{Item} & \multicolumn{2}{c}{User} \\
          & & \mlonem & \onion & \amazonvid & \mlonem & \onion & \amazonvid &  \mlonem & \onion \\\midrule
\multirow{2}{*}{\baserec} & \multirow{1}{*}{\rand} & 
0.0052 & 0.0009 & 0.0013 & 0.0503 & 0.0081 & 0.0110 & 0.0440 & 0.0087 \\
 & \multirow{1}{*}{\pop} & 
 0.1268 & 0.0182 & 0.0255 & 0.0087 & 0.0063 & 0.0210 & 0.4583 & 0.0994 \\
\midrule
\multirow{2}{*}{\cfbr} & \multirow{1}{*}{\mf~\cite{bpr}} & 
0.2589 & 0.1438 & \bfseries 0.0818$\dagger$ & 0.0300 & 0.0055 & 0.0125 & 0.4613 & 0.0920 \\
 & \multirow{1}{*}{\dmf~\cite{dmf}} & 
 0.1165 & 0.1297 & 0.0556 & 0.0097 & 0.0197 & 0.0210 & 0.4124 & 0.0575 \\
\midrule
\multirow{5}{*}{\cbr} & \multirow{1}{*}{\dropoutnetTwo~\cite{dropoutnet}} & 
0.2427 & 0.0749 & 0.0374 & 0.2032 & 0.1142 & 0.1033 & 0.4584 & 0.0947 \\
 & \multirow{1}{*}{\dropoutnetBest~\cite{dropoutnet}} & 
 0.2427 & 0.0935 & 0.0445 & 0.2032 & 0.1689 & 0.1033 & 0.4584 & 0.1086 \\
 & \multirow{1}{*}{\clcrec~\cite{clcrec}} & 
 \bfseries 0.2680$\dagger$ & 0.1378 & 0.0569 & 0.1762 & 0.1503 & 0.1348 & 0.4616 & 0.0989 \\
 & \multirow{1}{*}{\sbrecTwo} & 
 0.2593 & 0.1466 & 0.0728 & 0.2592 & 0.1457 & \bfseries 0.1479$\dagger$ & \bfseries 0.4673 & 0.0918 \\
 & \multirow{1}{*}{\sbrecBest} & 
 0.2561 & \bfseries 0.1616$\dagger$ & 0.0728 & \bfseries 0.2994$\dagger$ & \bfseries 0.1982$\dagger$ & \bfseries 0.1479$\dagger$ & 0.4659 & \bfseries 0.1094 \\
 
        \bottomrule
    \end{tabular}
    \vspace{0.6em}
    \caption{Evaluation results \wrt \ndcgten. Bold indicates the best performing RS, $\dagger$ indicates significant improvement over all others (paired $t$-tests with $p<0.05$ considering Bonferroni correction).}
    \label{tab:results-ndcg}
\end{table*}
\end{center}
In this section, we report the performance of \sbrec and of the baselines 
in standard and cold-start scenarios. We then analyze the effect of the single-branch in mapping multiple modalities to the shared embedding space. 
\subsection{Performance Comparison}
\begin{figure*}
    \centering
    \includegraphics[width=\textwidth]{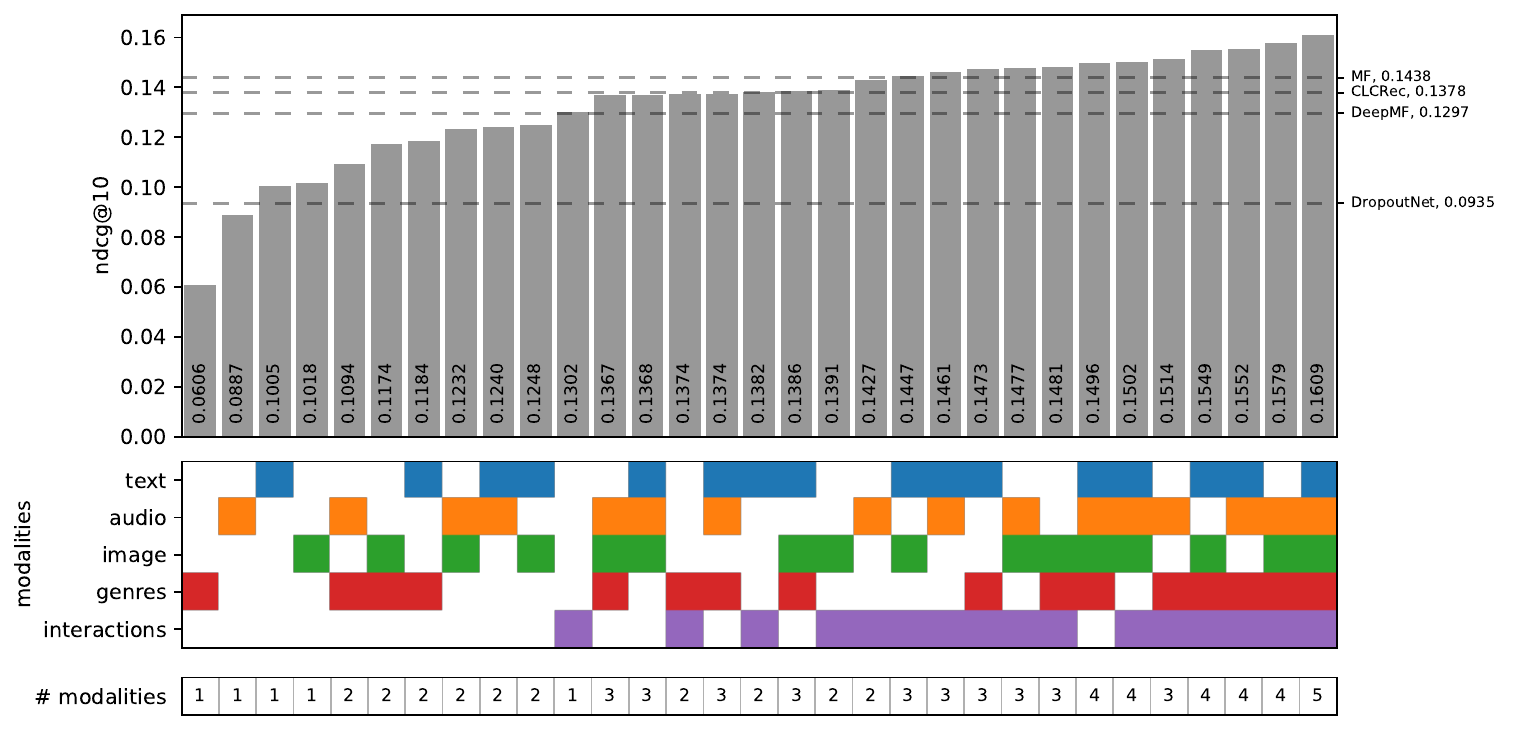}
    \caption{Performance of \sbrec on the test set of warm-start \onion, based on a varying set of modalities used. The bottom integers show the number of modalities. If a modality is used, its block is filled with the corresponding color in the central plot. The bar plot shows \sbrec's performance in terms of \ndcgten for each set of modalities. The gray dashed horizontal lines show the \ndcgten of \cfbr and \cbr algorithms.}
    \label{fig:missing-mod-ranking}
\end{figure*}

        \Tabref{tab:results-ndcg} shows the \ndcgten results\footnote{We refer the reader to the auxiliary material for similar tables reporting \precten, \recten, \apten and \covten.} of the algorithms on three datasets. For \onion and \mlonem, where both user and item information is available, we show the results on the three evaluation scenarios described in Section~\ref{sec:evaluation}. Since user information is not available for \amazonvid, we restrict ourselves to the random and item cold start for this dataset. Each row refers to a different algorithm. Solid lines divide the algorithms in the three categories described in Section~\ref{sec:evaluation}. For \sbrec (\dropoutnet), which allows leveraging more than one modality, we include both \sbrecTwo (\dropoutnetTwo) and \sbrecBest (\dropoutnetBest). The sign $\dagger$ indicates significant improvement of the best performing RS over all other RS (paired $t$-tests with $p<0.05$ considering Bonferroni correction to account for multiple comparisons).\\
    \textit{Warm split.}
        In the random split scenario, corresponding to the case in which all users and items in the test set are warm, overall \mf, \clcrec, and our \sbrec are among the best-performing algorithms, with the best result varying across recommendation domains. Specifically, \clcrec reaches the highest \ndcgten on \mlonem, slightly outperforming our \sbrecTwo and \mf . On  \onion, both versions of \sbrec outperform \mf and \clcrec. Finally, on the \amazonvid dataset, \mf outperforms \sbrec and \clcrec. \\
    \textit{Cold start.}
        The limitation of \cfbr techniques in cold-start scenarios is evident from the fact that they are outperformed by \cbrs. 
        This is more evident in the item than in the user cold-start scenario. We attribute the larger improvement of \cbrs with respect to \cfbr in the item cold start to the larger and more informative set of content modalities available for items; For cold-start users, where only demographic information is available, the difference between \cbrs and \cfbrs is therefore less pronounced. Our \sbrec outperforms all other algorithms (\cfbr and \cbrs) on all datasets and cold-start scenarios. The difference in performance is substantial when leveraging all the available modalities (\sbrecBest). This shows that by sharing the weights of the single-branch encoder, \sbrec is effectively leveraging the multimodality of the information available during the evaluation phase. \\
    \textit{Missing modalities.}
        As discussed in Section~\ref{sec:methodology}, the strength of \sbrec, which also makes it effective in cold-start scenarios, is its ability to tackle missing modality scenarios. It relies on the information extracted from the available modalities by means of a single branch, which is trained to encode information on the information shared by all modalities, including the missing one. \Figref{fig:missing-mod-ranking} highlights this by comparing the performance of \sbrecBest on \onion\footnote{We carry out this analysis on \onion since it is the dataset with the highest number of modalities available.} on the random split, corresponding to a warm-start scenario. The values and bars report the test \ndcgten reached by a single  \sbrec instance trained with all modalities, and evaluated leveraging different combinations of modalities. Since the model has been trained with five different modalities (interactions, audio, text, image, and item genre), $31$ different modality combinations are evaluated. Each modality corresponds to a different color. If it is used at evaluation, this is reflected by a colored box below the bar plot, while if it is not used, the box is left blank. The grey dashed lines show the \ndcgten reached by the comparison algorithms (\mf, \clcrec, \dmf, and \dropoutnet). Overall, the performance of \sbrec increases with an increasing number of modalities. 
        Moreover, considering the ranking of combinations of the same number of modalities, image is the most informative \newtext{side-information} modality, followed by text, audio, and genre. By only leveraging interaction data at evaluation provides better recommendations than leveraging any \newtext{side-information} modality or combinations of two of them. 
        In comparison to the other models, already leveraging only text or image content,
        \sbrec outperforms \dropoutnet, while either the simultaneous use of interaction and one content modality, or $3$ or more content modalities are required to outperform \dmf and \clcrec. Finally, leveraging at least $3$ modalities and including the interaction data allows \sbrec to outperform \mf.
    
    \subsection{Modality Gap}
\begin{figure}
    \centering
    \includegraphics[width=0.45\textwidth]{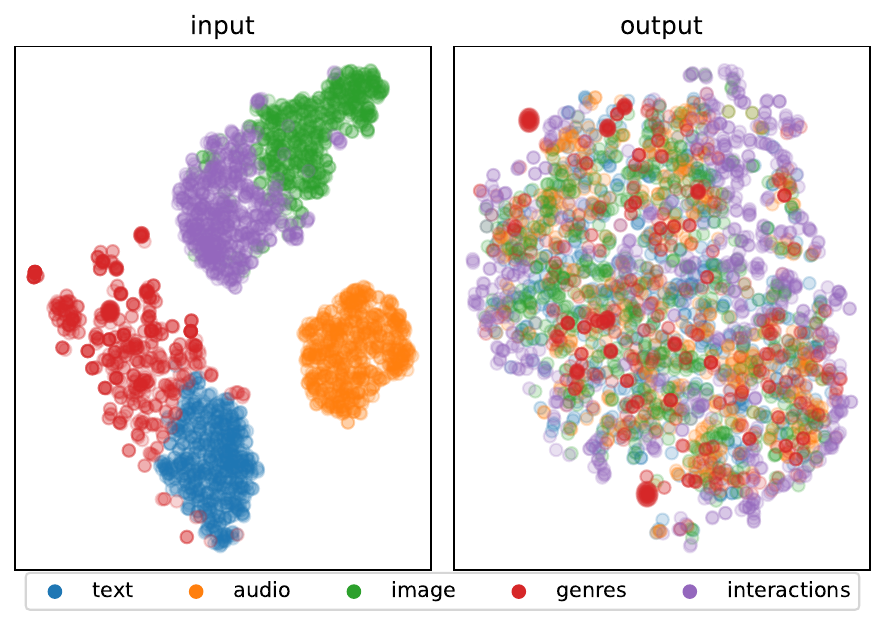}
    \vspace{-1em}
    \caption{t-SNE projected embeddings before and after \sbrec. While different modalities can be differentiated at \sbrec input (left), modalities overlap substantially in the shared embedding space after applying \sbrec (right).}
    \label{fig:projection-single}
\end{figure}

        In this section we analyze to which extent \sbrec is able to map different modalities to the same region of the embedding space, hence filling the modality gap often displayed by multimodal models~\cite{liang2022mindthegap,humer2023amumo}. We consider \sbrecBest trained on the \onion random split, which during training relies on all available modalities. 
        \Figref{fig:projection-single}
        shows the embeddings of the five item modalities (text, audio, image, genres, and interactions) after training.
        The left plot shows the embeddings used as input to the single-branch network \newtext{$g$}, while the right one shows the output of the single-branch network.  To visualize the high-dimensional vectors in two dimensions, we first select the first $10$ principal components through principal component analysis. We then apply T-distributed Stochastic Neighbor Embedding (t-SNE)~\cite{hinton2002tnse}\footnote{We use t-SNE with the default parameters reported by the library scikit-learn~\cite{sklearn_api} and refer the reader to its documentation for details, (\url{https://scikit-learn.org/stable/modules/generated/sklearn.manifold.TSNE.html}).} to further reduce the dimensions to $2$. To improve clarity, we only show the embeddings of a subset of all items, consisting of  $3{,}000$ items sampled uniformly at random.
        The left plot of \Figref{fig:projection-single} shows that before the single-branch network, embeddings of different modalities cover different regions of the embedding space. On the contrary,
        the right plot shows that the single-branch network maps all the modalities to the same region of the embedding space. This is an indication of \sbrec's ability to extract representations that are similar -- and therefore similarly effective for recommendation -- from any of the item modalities.
        
\section{Conclusion and Future Work}
    \label{sec:conclusions}
    In this paper, we propose \sbrec, a novel multimodal RS that leverages a single-branch architecture to encode multimodal user and item information, including collaborative data. The single-branch design allows \sbrec to provide accurate recommendations from any modality. As a result, \sbrec's recommendations are accurate also in scenarios of missing modality, including cold start. We show through extensive quantitative experiments that \sbrec significantly outperforms \cfbr as well as \cbrs in terms of accuracy of recommendations in item cold-start scenarios, and that it is competitive with both \cfbr and \cbrs in warm-start as well as user cold-start scenarios. Furthermore, we analyze the impact of missing modalities on the performance of \sbrec, showing that the proposed model reaches its best performance when all modalities are leveraged, but it is still able to outperform \cfbr and other \cbrs models when input modalities are missing. Moreover, we analyze \sbrec's embedding space shared by the multiple modalities and show that as a result of its design, \sbrec is able to reduce the modality gap. \sbrec's ability to combine multimodal representations relies on the use of the single-branch, which does not exclude the use of a contrastive loss. The impact of the contrastive loss on \sbrec 
    can be analyzed by means of an ablation study;
    \newtext{ The performance of \sbrec can also be compared with that of multi-branch architectures with contrastive loss. We focused on a version of \sbrec with an underlying RS architecture similar to \dmf and envision as future work variants of \sbrec based on other core RS. A deeper analysis on the modality gap of \sbrec 
    could analyze the distance in the embedding space between multimodal embeddings of a same user or item, as well as the correlation between modality gap and model performance. Our analysis of missing-modality scenarios was limited to the case in which all modalities are available during training but not during inference, \ie we did not consider missing modalities at training time.
    Additionally, our work did not consider the scalability of \sbrec in function of the dimensionality of the input modalities. {Although this scenario is partially addressed in the analysis summarized in Figure~\ref{fig:missing-mod-ranking}, we did not evaluate the performance of RSs in scenarios where all items in the validation and test sets have no information regarding one of the modalities related to side information. }} Finally, reducing the amount of collaborative data while increasing content information might substantially impact the performance of \sbrec in terms of beyond-accuracy metrics. 
    We leave these extensions of the current work for future research.

\begin{acks}
    This research was funded in whole or in part by the Austrian Science Fund (FWF) \url{https://doi.org/10.55776/P33526}, \url{https://doi.org/10.55776/DFH23}, \url{https://doi.org/10.55776/COE12}, \url{https://doi.org/10.55776/P36413}. \newtext{For open access purposes, the authors have applied a CC BY public copyright license to any author-accepted manuscript version arising from this submission.
    
     Christian Gahnör and Marta Moscati would like to thank Alessandro Melchiorre for his invaluable work in developing the initial version of the framework used to benchmark \sbrec with other RSs.}
\end{acks}

\clearpage

\bibliographystyle{ACM-Reference-Format}
\FloatBarrier\balance\bibliography{main.bib}

\end{document}